# Magnetic, magnetocaloric and magnetotransport properties of $RSn_{1+x}Ge_{1-x}$ compounds (R=Gd, Tb, Er; x=0.1)

Sachin B. Gupta[1], K.G. Suresh[1*] and A.K. Nigam[2]

[1]*Department of Physics, Indian Institute of Technology, Mumbai-400076, India*

[2]*Tata Institute of Fundamental Research, Homi Bhabha Road Mumbai-400005, India*

## Abstract

We have studied the magnetic, magnetocaloric and magnetotransport properties of $RSn_{1+x}Ge_{1-x}$(R=Gd, Tb, Er; x=0.1) series by means of magnetization, heat capacity and resistivity measurements. It has been found that all the compounds crystallize in the orthorhombic crystal structure described by the centrosymmetric space group Cmcm (No. 63). The magnetic susceptibility and heat capacity data suggest that all the compounds are antiferromagnetic. Large negative values of $\theta_p$ in case of $GdSn_{1.1}Ge_{0.9}$ and $TbSn_{1.1}Ge_{0.9}$ indicate that strong antiferromagnetic interactions are involved, which is also reflected in the magnetization isotherms. On the other hand $ErSn_{1.1}Ge_{0.9}$ shows weak antiferromagnetic interaction. The heat capacity data have been analyzed by fitting the temperature dependence and the values of $\theta_D$ and $\gamma$ have been estimated. Among these three compounds, $ErSn_{1.1}Ge_{0.9}$ shows considerable magnetic entropy change of 9.5 J/kg K and an adiabatic temperature change of 3.2 K for a field of 50 kOe. The resistivity data in different temperature regimes have been analyzed and the dominant contributions have been identified. All the compounds show small but positive magnetoresistance.

*Corresponding author (email: suresh@phy.iitb.ac.in)

# I. Introduction

The rare earth (R) intermetallics have been extensively studied for many years due to their greatly varying magnetic and electrical properties. These intermetallics exist in different compositions, with magnetic as well as nonmagnetic transition metal constituents. One of the interesting classes in this series is R-T-X where T is a transition metal and X is a p block element. Recently, nearly equiatomic compounds with general formula $RSn_{1+x}Ge_{1-x}$ (R=Y,Gd-Tm, x=0.15) have been reported by Tobash *et al.* [1]. The crystal structure of these compounds is similar to that of $RT_xX_2$ compounds [2,3]. In the unit cell of these compounds, all the atoms are located at 4c site (0, y, 1/4) with different values of positional parameter $y_i$ for the constituent elements. All the compounds order antiferromagnetically in the temperature range between 4 and 30 K. The end members of this series, *viz.* $RSn_2$ (R=Tb, Dy) [4] have a Neel temperature ($T_N$) of 22, 15 K respectively [4]. $RGe_2$ compounds are also antiferromagnetic with $T_N$ of 28, 42, 28 and 11 K for $GdGe_2$, $TbGe_2$, $DyGe_2$ and $HoGe_2$ respectively [5]. Therefore it is clear that substitution of Ge for Sn results in an increase in the Neel temperature in this series. The magnetic structure of $RSn_{1+x}Ge_{1-x}$ (R=Tb, Er) compounds with x= 0.12 and 0.08 has been investigated by neutron diffraction study [6]. These studies have revealed that the magnetic moment is localized only on the rare earths atoms, which suggests that magnetic properties in general are governed by the RKKY exchange interaction and the crystal field effects.

Recently, there is an intense search for novel and potential materials for magnetocaloric applications and in this respect a large number of intermetallic compounds are being investigated. This has also led to a better understanding of the magnetism in such materials. Though the magnetic and transport studies have been reported in RSnGe series in some detail, there are no reports on the magnetocaloric effect (MCE). Therefore, in this work we report a detailed study of $RSn_{1+x}Ge_{1-x}$ (R=Gd, Tb, Er; x=0.1) compounds with the intention of correlating the magnetocaloric properties with the basic magnetic properties. We have also investigated the transport properties of this series. For the sake of simplicity, the series $RSn_{1.1}Ge_{0.9}$ is referred as RSnGe throughout this paper.

## II. Experimental Details

The polycrystalline samples were synthesized by arc melting the stoichiometric amounts of the constituent elements of high purity (99.9% for rare earths and 99.99% for Sn and Ge). The melted ingots were flipped and melted four times. They were then sealed in quartz tube in a vacuum of $10^{-6}$ torr and annealed for a week at 500 ˚C to ensure the homogeneity.

The X-ray powder diffraction (XRD) data were collected at room temperature on a X'PERT PRO using Cu Kα radiation. DC magnetization measurements, M (T) and M (H), were performed using a Vibrating Sample Magnetometer (VSM) attached to a Physical Property Measurement System (Quantum Design, PPMS-6500). The M-T data were recorded from 2 to 300 K under zero field (ZFC) and field cooled warming (FCW) modes. In the ZFC mode the sample was cooled in the absence of field from 300 K to 2 K, then a constant field was applied and the magnetization was measured with increasing temperature. In the FCW mode the sample was cooled from the paramagnetic region to the measurement temperature in presence of a field, but the data were recorded only during warming.

The heat capacity measurements were performed with and without field in the temperature range from 2-100 K in PPMS using thermal relaxation technique. The electrical resistivity was measured in zero field as well as in presence of a field of 50 kOe in the temperature range of 2-300 K using the linear four probe method, applying an excitation current of 150 mA. The bar shaped samples for the electrical resistivity measurement were cut from the annealed ingot by diamond saw. The dimensions of the samples were close to 1x1x7 mm.

## III. Results and Discussion

The XRD pattern (Fig. 1) confirms that all the compounds crystallize in $ZrSi_2$ type orthorhombic structure with centrosymmetric space group Cmcm (No. 63) with traces of impurities in some of them. The crystal structure and other parameters of all the compounds were calculated by Rietveld refinement. The lattice parameters are found to be a= 4.301(6)Å, b=16.484(2)Å, c=4.094(6)Å for GdSnGe and a=4.269(3)Å, b=16.333(1)Å, c=4.061(3)Å for TbSnGe and a=4.224(3)Å, b=16.061(1)Å, c=4.039(3)Å for ErSnGe, all in good agreement with reported values of compounds with comparable x value [1].

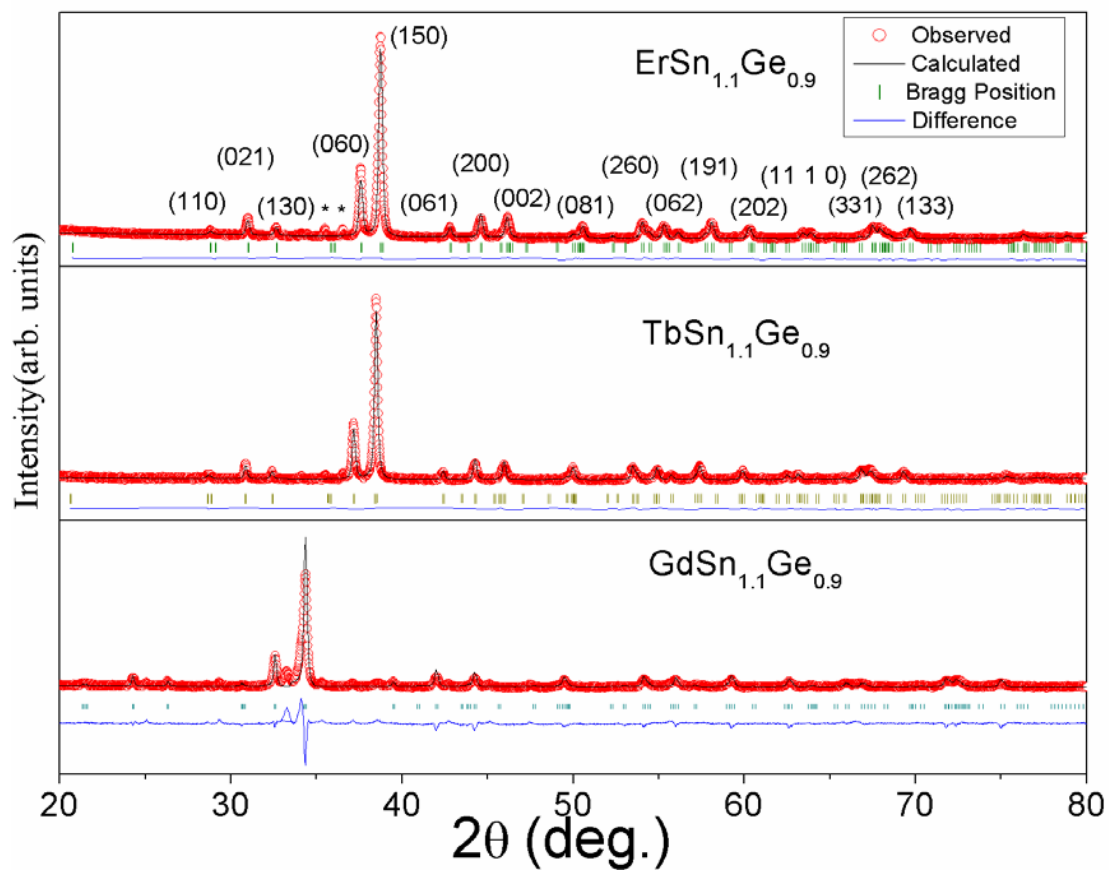

Fig. 1. Powder X-ray diffraction patterns, along with the Rietveld refinement of RSnGe (R=Gd, Tb, Er) compounds. The plots at the bottom show difference between the theoretical and the experimental data.

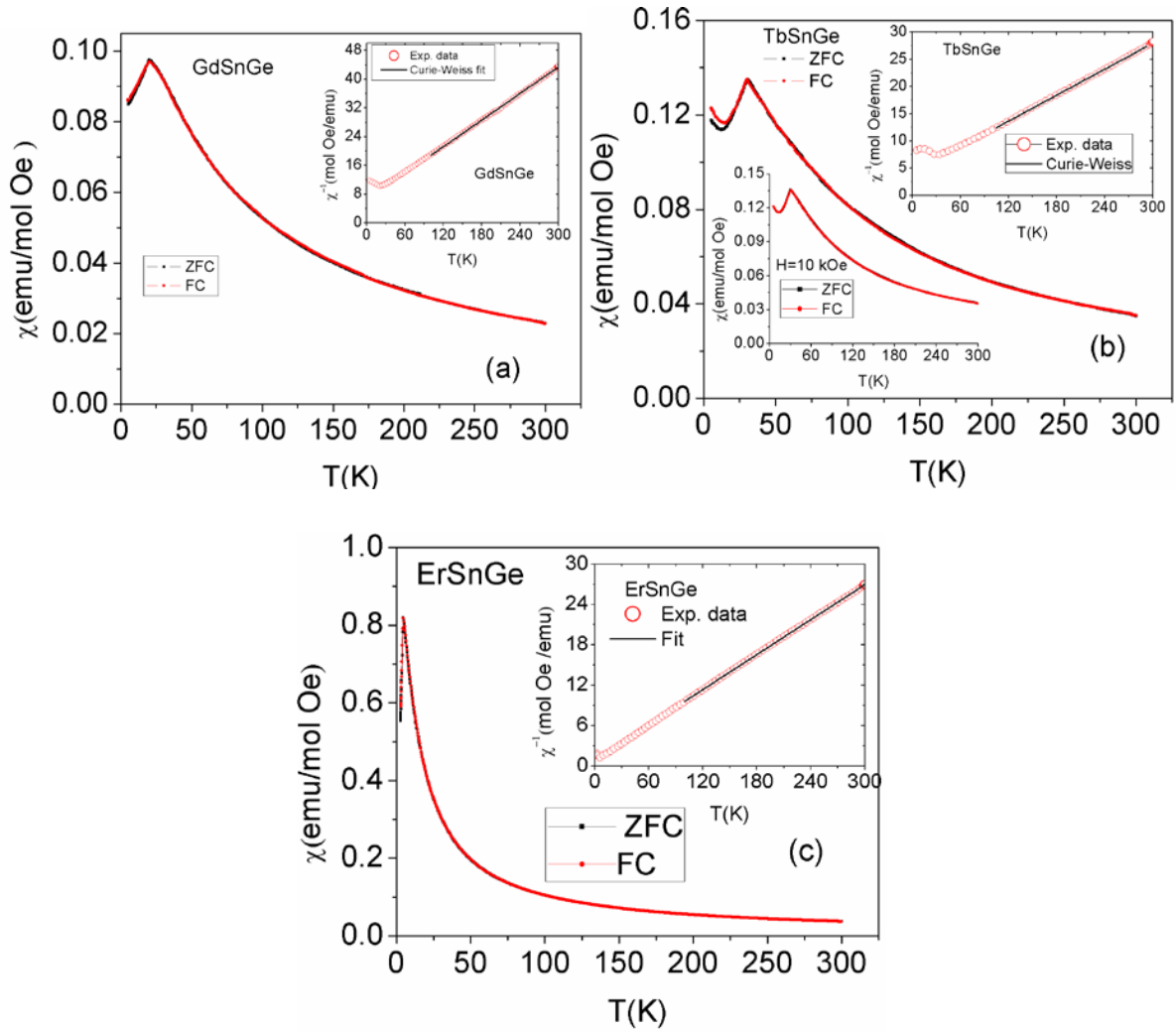

Fig.2. Temperature (T) dependence of magnetic susceptibility (χ) for RSnGe (R=Gd,Tb,Er) compounds collected under ZFC and FCW modes. The insets show the temperature variation of inverse susceptibility obtained in H=500 Oe data along with the Curie-Weiss fit. The lower inset in (b) shows the effect of 10 kOe field on the thermomagnetic irreversibility.

Temperature dependence of magnetization of all the compounds obtained under various fields, both under ZFC and FCW conditions are shown in figure 2. The susceptibility (χ) vs. T plots of these compounds, which are obtained in a field of 500 Oe are shown in Figs.2(a)-2(c). It is clear from these plots that all the compounds are antiferromagnetic at low temperatures. The Neel temperatures for these compounds, which are obtained by plotting the first derivative of magnetization with respect to temperature (dM/dT), are given in table I.

Among all compounds, TbSnGe is found to have the highest Neel temperature (30 K). The dc magnetic susceptibility for all the compounds above 100 K obeys the Curie –Weiss law. The effective magnetic moment ($\mu_{eff}$) and the paramagnetic Curie temperature ($\theta_p$) for all the compounds are obtained from the Curie-Weiss fit (see upper insets of Fig. 2) using the relations, $\chi^{-1} = (T - \theta_p)/C_m$ and $\mu_{eff} = (8C_m)^{1/2}$ (here $C_m$ is molar Curie constant). It is evident from table I that the paramagnetic moments calculated from the Curie-Weiss fit agree well with theoretical values of $R^{+3}$ free ions as well as reported values of compounds with x=0.15 [1]. The negative value of $\theta_p$ is consistent with the antiferromagnetic ordering. Large negative values of $\theta_p$ in the case of Gd (-49K) and Tb (-37K) compounds indicate that strong magnetic interactions are present in these two compounds. On the other hand, ErSnGe has a relatively smaller $\theta_p$ value (-11 K).

Table I. Values of Neel temperature ($T_N$), paramagnetic Curie temperature ($\theta_p$), effective magnetic moment observed ($\mu_{eff}$) and calculated ($g\sqrt{J(J+1)}$), maximum entropy change ($\Delta S_M^{Max}$) and max. adiabatic temperature change ($\Delta T_{ad}^{Max}$) close to the Neel temperature in RSnGe compounds.

| Compound | $T_N$ (K) | $\theta_p$ (K) | $\mu_{eff}$ ( $\mu_B/R^{3+}$) | $g\sqrt{J(J+1)}$ ( $\mu_B/R^{3+}$) | $-\Delta S_M^{Max}$ (J/kg K) | $\Delta T_{ad}^{Max}$ (K) |
|---|---|---|---|---|---|---|
| GdSnGe | 20 | -49 | 8.04 | 7.94 | -- | -- |
| TbSnGe | 30 | -37 | 9.8 | 9.72 | -0.7 | -0.27 |
| ErSnGe | 4.7 | -11 | 9.6 | 9.6 | 9.5 | 3.2 |

I

In TbSnGe, below 12 K, there is an upturn in magnetization data (see Fig.2b), which may be due to a spin reorientation. This also corroborates with the reported neutron diffraction data on compounds with comparable x values by Gil et al. [6]. These authors have reported that the magnetic structure of TbSn$_{1.12}$Ge$_{0.88}$ is sine wave modulated and is different from other members of the series. At 1.62 K, in this compound, all the Tb moments are aligned in same direction and are coupled ferromagnetically. Therefore, a similar scenario may be present in the present TbSnGe compound as well. Furthermore, from magnetization isotherms (shown in Fig. 4b and discussed later), one can see that all the isotherms increase linearly with field above 5 K, but at 5 K there is a change in the slope, which also indicates the spin reorientation. Furthermore, it can be seen that among these three compounds, TbSnGe exhibits the largest thermomagnetic irreversibility (between ZFC and FCW data), which is attributed to the pinning of domain walls [7-9]. It is well known that the width of the domain wall in a ferromagnet is directly proportional to the magnetic ordering temperature ($T_{ord}$) as given by the equation

$$\delta = \sqrt{\frac{s^2 \pi^2 J_{ex}}{aK}}; \quad i.e., \delta \propto \sqrt{\frac{T_{ord}}{K}} \tag{1}$$

Thus systems with low ordering temperatures have relatively narrow domain walls. In some cases, it was seen that the systems with low transition temperatures (narrow domain walls) show considerable pinning effect [10] at low temperatures. In the ZFC mode, when the sample is cooled from paramagnetic region, the domain walls become frozen (pinned) and less mobile which reduces the magnetization at low temperature. On increasing the temperature, the mobility of the domain walls increases due to the increase in the thermal energy and hence the magnetization increases. It is also observed that the domain wall pinning effect is generally seen in materials with large magnetic anisotropy [10]. It may also be noted from Fig. 2(b) that the thermomagnetic irreversibility completely disappears at 10 kOe.

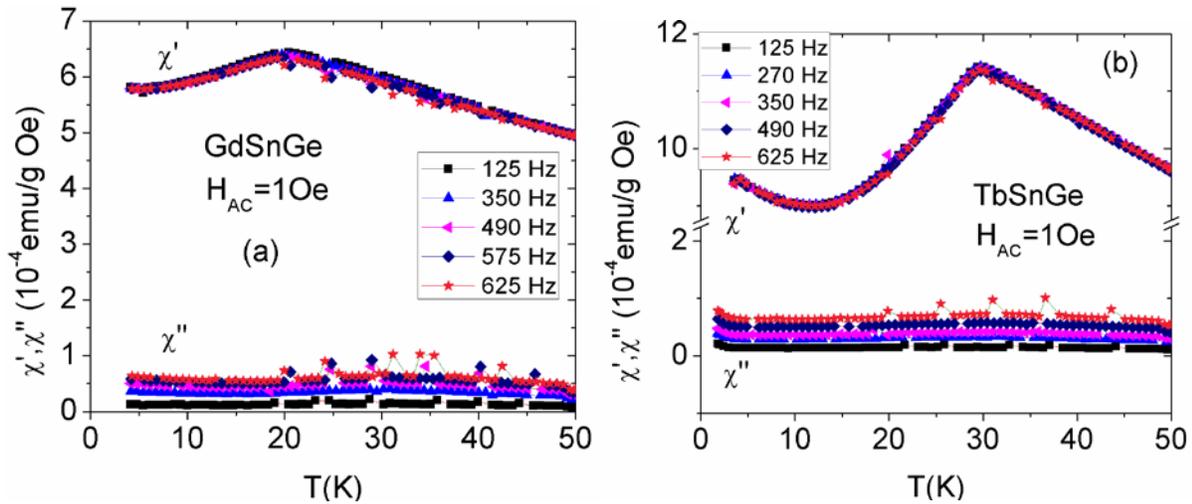

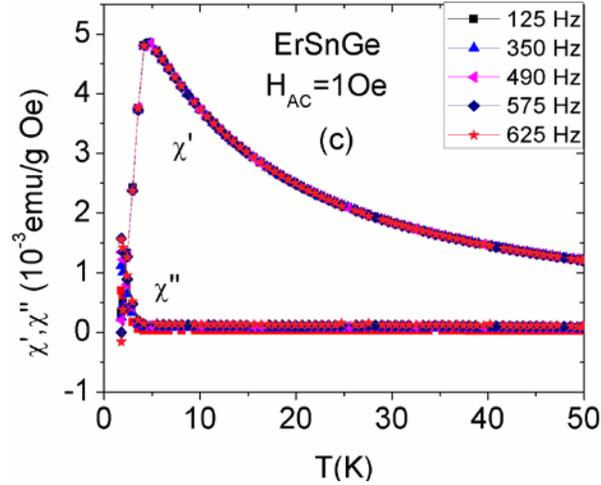

Fig. 3. Zero field ac magnetic susceptibility data at an oscillating field amplitude of 1Oe recorded at different frequencies.

AC susceptibility (ACS) measurement was performed in ZFC condition in the temperature range 1.8-50 K. Zero field ACS is one of the best methods to analyze the magnetic transitions and the associated magnetization dynamics since the oscillating magnetic field does not disturb the system very much. The in phase and out of phase susceptibilities, $\chi'(T)$ and $\chi''(T)$, were recorded simultaneously at different frequencies and at constant oscillating field amplitude ($H_{AC}$) of 1 Oe for all the compounds, as shown in Fig. 3. GdSnGe and TbSnGe show a peak in $\chi'(T)$, corresponding to the antiferromagnetic transition. No peak is observed in the entire temperature range investigated for the $\chi'(T)$, thereby suggesting that the transition is indeed antiferromagnetic. On the other hand, ErSnGe shows a peak in both $\chi'(T)$ and $\chi''(T)$, with a small shift in the peak positions. Therefore, it is quite clear that the magnetic state in ErSnGe is somewhat different from that of the other two compounds.

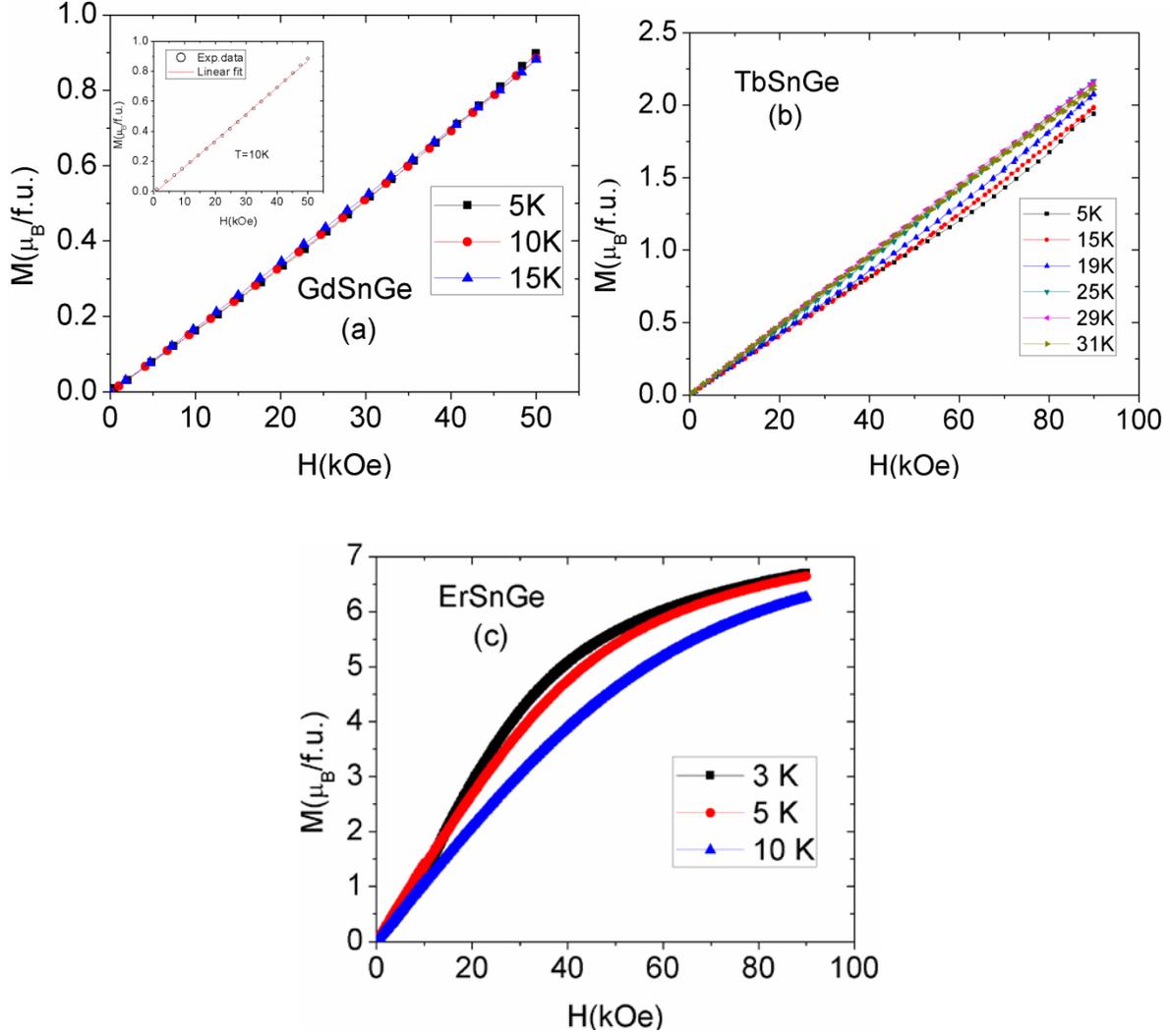

Fig.4. M vs. H curves for (a) GdSnGe, (b) TbSnGe and (c) ErSnGe obtained at selected temperatures for fields upto 90 kOe. Upper inset in (a) shows the linear fit to M-H data.

The magnetization isotherms collected at various temperatures below and above $T_N$ and in fields up to 90 kOe for RSnGe are plotted in Fig.4. The magnetization data in all compounds do not show any hysteresis. The inset in Fig. 4(a) shows the linear fit to magnetization data for GdSnGe. In case of TbSnGe, magnetization data (except at 5K) show the linear dependence with the field. This linear dependence of magnetization in GdSnGe and TbSnGe is due to the strong antiferromagnetic correlation. However, in TbSnGe, at 5 K, the plot deviates from linearity at high fields, indicating the presence of weak ferromagnetic coupling of moments. It is also of interest to note from figure 4(b) that below $T_N$ magnetization increases as temperature increases even for a field as high as 90 kOe, which also proves the strong antiferromagnetic behavior of TbSnGe compound. On the other hand, above $T_N$, the magnetization tends to decrease with increase in temperature. Interestingly, ErSnGe shows a slightly different behavior. In the low field region, the magnetization at 5 K is more than that at 3 K. But at higher fields, the trend reverses, which indicates that the antiferromagnetic coupling is broken at a field of about 14 kOe and the moments try to couple ferromagnetically, which is called the spin flop transition. It is also worth noting here that the isothermal magnetization data do not show any indication of saturation in GdSnGe and TbSnGe, while in ErSnGe, it shows the saturation trend at fields close to 90 kOe. The curvature seen at all the three temperatures in ErSnGe is of importance as it does not fit into a strong antiferromagnetic behavior.

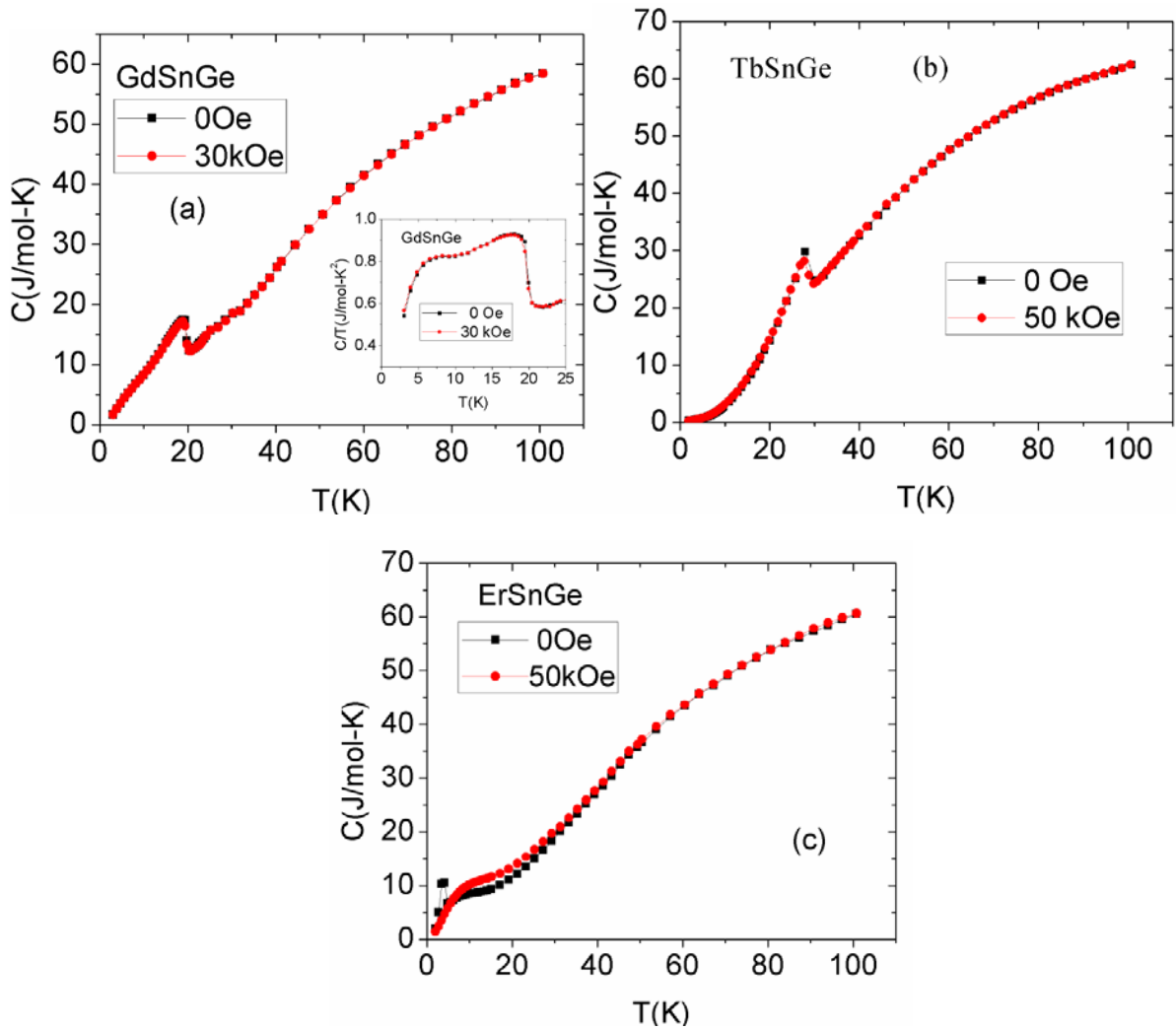

Fig. 5. Temperature variation of heat capacity in different fields for (a) GdSnGe (b) TbSnGe and (c) ErSnGe.

In order to further probe the nature of the magnetic behavior of these compounds, the heat capacity measurements were carried out under zero field and in various fields in the temperature range of 2-100 K and the data are shown in Fig. 5. It is well known that the heat capacity in metallic, magnetic materials can be considered as the sum of electronic, lattice and magnetic contributions, as given follows

$$C(T) = C_{el}(T) + C_{lat}(T) + C_{mag}(T) \qquad (2)$$

The electronic and lattice contributions (nonmagnetic) is given by the equation [11]

$$C_{nonmagnetic} = C_{el} + C_{lat} = \gamma T + 9NR(T/\theta_D)^3 \int_0^{\theta_D/T} \frac{x^4 e^x}{(e^x - 1)^2} dx \qquad (3)$$

Here, N is the number of atoms per formula unit, R is the molar gas constant, $\gamma$ is the electronic coefficient and $\theta_D$ is the Debye temperature. A reasonable estimate of the values of $\gamma$ and $\theta_D$ has been obtained by fitting of equation (3) to the total heat capacity C(T) with a single Debye temperature. In the case of GdSnGe, the fit yielded $\theta_D$ =290 K and $\gamma$ =80 mJ/mol K$^2$.

The inset in Fig. 5(a) shows the C/T vs. T plot for GdSnGe in the expanded scale, which shows two humps one at 20 K, which corresponds to the antiferromagnetic ordering. The other one is attributed to the spin reorientation [1]. It may be noted that the signature of second hump is not seen in magnetization measurements.

Fig. 5 also shows that all the compounds show $\lambda$- singularity at temperature close to their $T_N$, implying that the magnetic transition is of second order in nature. It can also be seen that, on applying a field, the peak gets suppressed and shifts to slightly lower temperatures, once again confirming the antiferromagnetic nature of the transition. In ErSnGe, on the application of field, the peak becomes broader. This lends another evidence for the weak antiferromagnetic coupling in this compound. The peak broadening induced by applied field is usually seen in ferromagnets due to the modification of the ordering temperature, which causes a suppression of the peak.

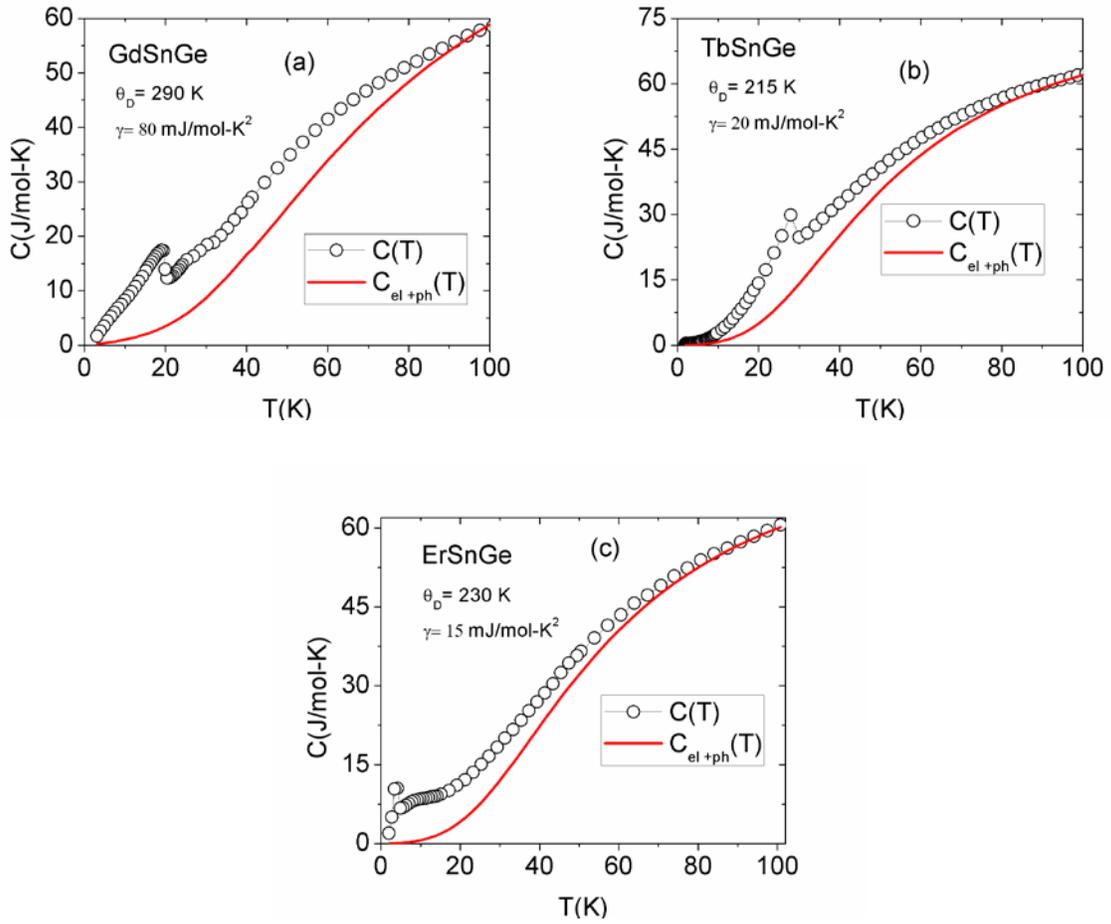

Fig. 6. Fit of the nonmagnetic contribution to the heat capacity in RSnGe compounds.

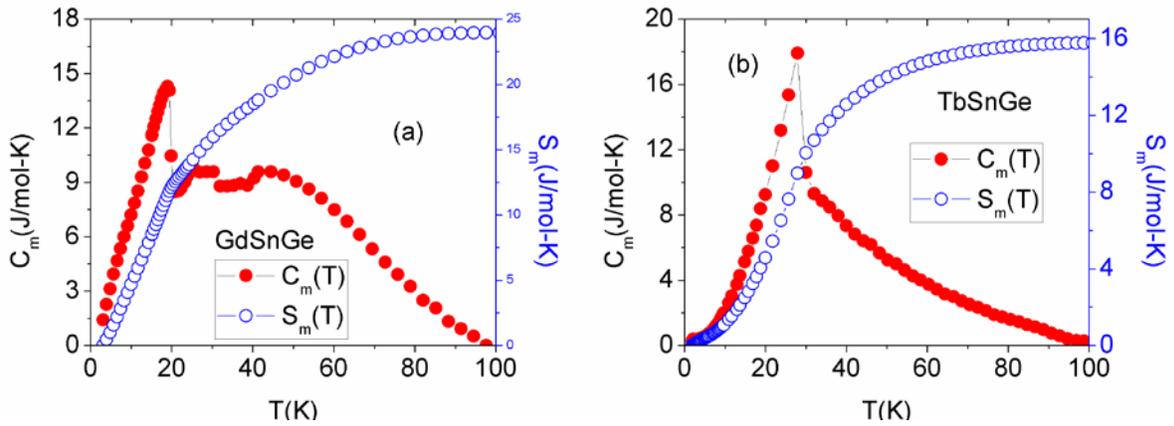

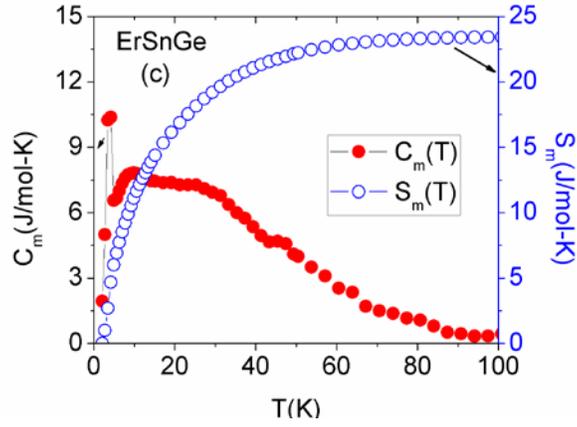

Fig.7. Variation of magnetic heat capacity and the magnetic entropy with temperature in RSnGe compounds.

The magnetic contribution to the heat capacity of RSnGe has been estimated by subtracting the non magnetic part ($C_{el+ph}$) from total heat capacity (C). The fit of the non magnetic part is shown in Fig. (6) and the resulting magnetic contribution is plotted in Fig. 7. A broad hump above $T_N$ in the magnetic heat capacity of GdSnGe and ErSnGe is found, which can be attributed to Schottky anomaly. The magnetic entropy is calculated numerically by integrating the magnetic heat capacity data, $S_m = \int_0^T \frac{C_m}{T} dT$. The total entropy obtained at 100 K for TbSnGe is 15.7 J/mol K, which is less than the expected value R ln(2J+1) for the $Tb^{3+}$. This low value signifies that crystal field effect is present in this sample. In case of ErSnGe, the total entropy obtained at 100 K is 23.2 J/mol K, which is close to expected value of R ln(2J+1) for $Er^{3+}$. In GdSnGe, the estimated $S_m$ value is higher than the R ln(2J+1) for the $Gd^{3+}$, the reason for which is not understood now.

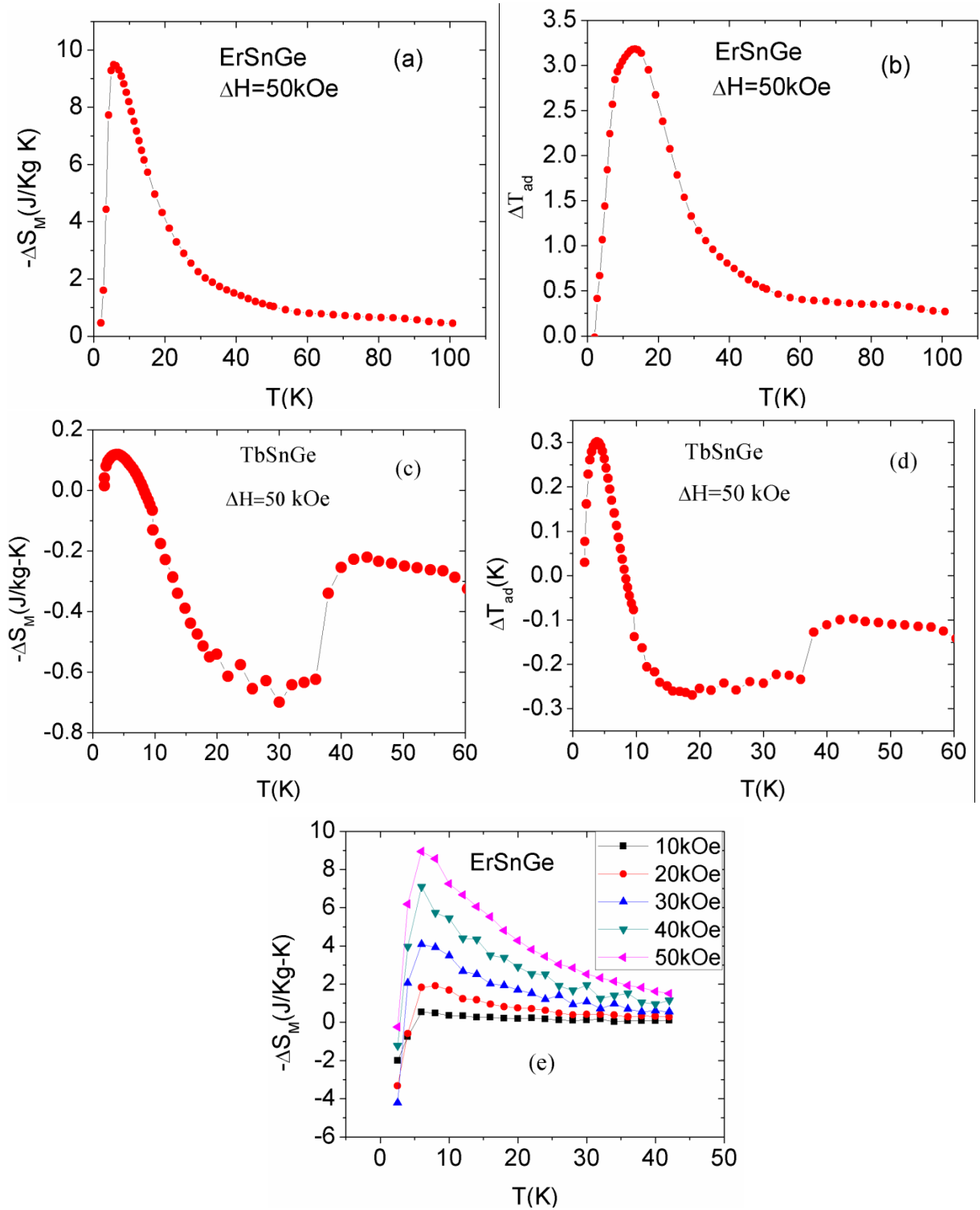

Fig. 8. (a,c) $\Delta S_M$ vs. T plot for ErSnGe and TbSnGe, (b,d) $\Delta T_{ad}$ vs. T plot for ErSnGe and TbSnGe (e) $\Delta S_M$ vs. T plot at different fields for ErSnGe calculated from the M-H-T data.

Magnetization and the heat capacity data have been independently used to calculate the magnetocaloric effect of RSnGe (R=Tb, Er) compounds. The magnetization data obtained at temperatures close to $T_N$ have been used to calculate MCE using Maxwell's relation, $\Delta S_M = \int_0^H \left[ \frac{\partial M}{\partial T} \right]_H dH$ [12]. Further, MCE in terms of isothermal magnetic entropy change ($\Delta S_M$) and adiabatic temperature change ($\Delta T_{ad}$) have been estimated from the heat capacity data utilizing the following thermodynamic relations [13]:

$$\Delta S_M(T,H) = \int_0^T \frac{C(T',H) - C(T',0)}{T'} dT' \qquad (4)$$

$$\Delta T_{ad}(T)_{\Delta H} \cong \left[ T(S)_{H_f} - T(S)_{H_i} \right]_S \qquad (5)$$

Figs. 8(a) and 8(c) show the temperature variation of isothermal entropy change ($\Delta S_M$) calculated from heat capacity data for ErSnGe and TbSnGe respectively. It can be seen from these figures that $\Delta S_M$ vs. T plots of both compounds have a maxima near the Neel temperature. The maximum values of $\Delta S_M$ for a field change ($\Delta H$) of 50 kOe in ErSnGe and TbSnGe compounds are 9.5 J/kg K and -0.7 J/kg K respectively. Fig. 8(e) shows the temperature variation of $\Delta S_M$ for ErSnGe (for different field changes) calculated from the M-H-T data. It may be noticed from this figure that $\Delta S_M$ value calculated from the *C-H-T* data compare very well with that calculated using the *M-H-T* data. The MCE in terms of adiabatic temperature change ($\Delta T_{ad}$) is shown in Figs. 8(b) and 8(d) for Er and Tb compounds respectively. The maximum values in this case are found to be -0.27 and 3.2 K for TbSnGe and ErSnGe respectively. It is worth noting from Fig. 8(c) that the entropy change is negative (positive MCE) below ~12K in the case of TbSnGe and it changes sign (becoming positive below (negative MCE)) at higher temperatures. A similar behavior can also be seen from $\Delta T_{ad}$ vs. T plot (see fig. 8(d)). It is well known that MCE of ferromagnetic materials are positive and negative for antiferromagnetic materials. Thus, the MCE data confirms that there is some weak ferromagnetic coupling of moments below 12 K in TbSnGe. Another important observation is that ErSnGe shows negative entropy change throughout the temperature regime of investigation, which proves that the antiferromagnetic interactions in this compound are rather weak. It is of interest to mention that the $\Delta S_M^{max}$ value of Gd$_2$PdSi$_3$ [14], which is proposed as a potential magnetic refrigerant below 40K is 8 J/Kg K ($\Delta$H=40 kOe). The maximum values of $\Delta S_M$ ($\Delta S_M{}^{max}$) of other promising magnetic refrigerants such as (Er,Dy)Al$_2$ vary in the range of 20–35 J/ kg K for $\Delta H$=50 kOe [15]. The compounds GdNiAl and GdPdAl [16] have $\Delta S_M{}^{max}$ values of 10.6 and 10.4 J/kg K for $\Delta H$=50 kOe. These facts suggest that ErSnGe possesses MCE values comparable to many known systems. Since the magnetic moment in the case of GdSnGe is very small compared to ErSnGe and TbSnGe, very low value of entropy change has been obtained in GdSnGe. The MCE parameters along with the magnetization data are listed in Table I.

The value of relative cooling power (RCP) is calculated by integrating area under the $\Delta S_M$ vs. T curve, using the relation $RCP = \int_{T_1}^{T_2} \Delta S_M(T) dT$, where T$_1$ and T$_2$ are the temperature limits at half maximum of $\Delta S_M$. Physically, the RCP is the measure of heat transfer from the cold reservoir to the hot reservoir in an ideal refrigeration cycle. The RCP value calculated from C-H-T data of ErSnGe is 67 J/kg, which is also comparable to that of some potential materials.

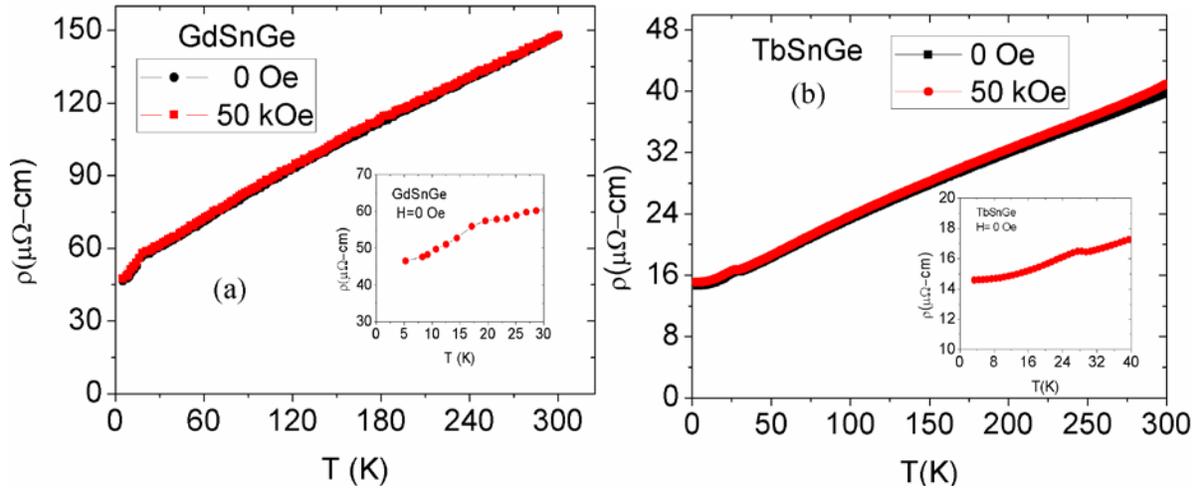

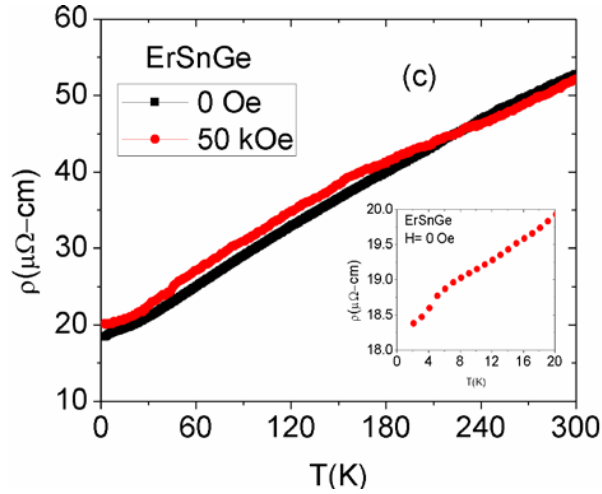

Fig. 9. Temperature variation of electrical resistivity in zero and 50 kOe for (a) GdSnGe (b) TbSnGe and (c) ErSnGe. The insets show zero field resistivity data in expanded scale.

The electrical resistivity is very sensitive to magnetic nature of materials and hence, the electrical resistivity data reflects the change in magnetic state. Thus, to further understand the magnetic behavior of RSnGe system, we have measured the temperature dependence of electrical resistance under zero and 50 kOe fields and the results are shown in figure 9. Insets in Fig.(9) give a clear view of resistivity data at low temperatures for all the compounds. It can be seen from this figure that in all the compounds, the temperature dependence of resistivity is linear above the Neel temperature, as expected for metallic systems. At around $T_N$, GdSnGe shows a sharp drop in resistivity followed by linear decrease with decrease in temperature. TbSnGe shows slightly different behavior around $T_N$. A drop in resistivity at the magnetic transition is expected because of the loss of spin-disorder contribution but there is a slight increase in resistivity associated with Neel temperature. Generally, increase in resistivity around antiferromagnetic ordering is attributed to two reasons: (i) carrier scattering by critical spin fluctuation (ii) formation of an antiferromagnetic super zone gap due to the additional periodicity of antiferromagnetic state [17]. Therefore, the increase seen in TbSnGe suggests that the effect due to the formation of magnetic Brillouin-zone boundary gap is larger in this case. A similar effect has also been observed in some other intermetallic compounds such as $GdPt_2Ge_2$, $Dy_7Rh_3$ and $PrCu_2Si_2$ [18-20].

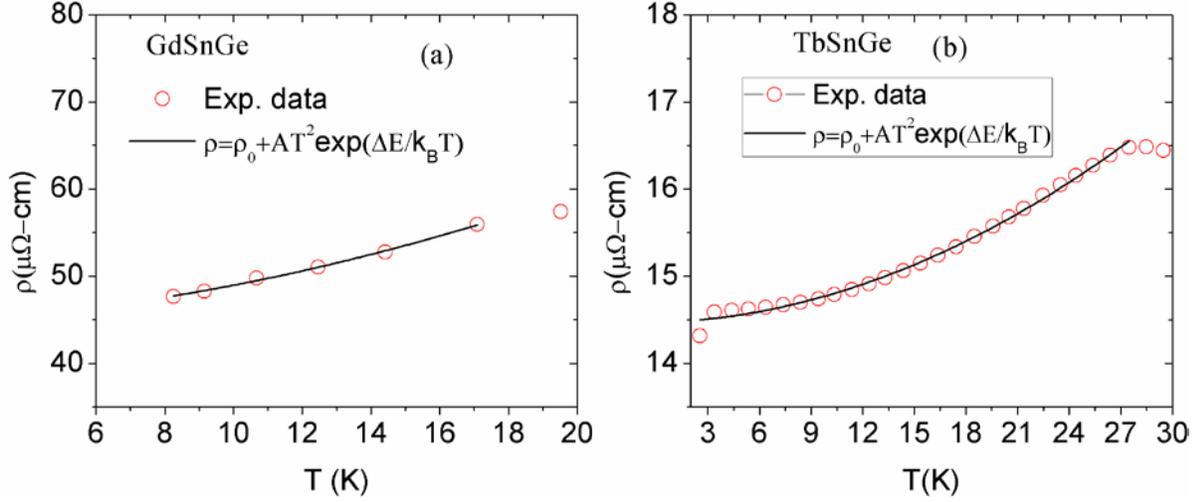

Fig. 10. Fit of equation $\rho(T) = \rho_0 + AT^n \exp(\Delta E/k_B T)$ to zero field resistivity data of (a) GdSnGe and (b) TbSnGe below transition temperature.

It is well known that the temperature dependent part of resistivity at low temperatures in magnetic materials is mainly due to (a) electron-electron scattering [21] (b) electron-spin wave scattering [21,22], and (c) electron-phonon scattering [23].These dependencies are usually different for ferromagnets and antiferromagnets. In general, at very low temperatures, the resistivity is mainly contributed by the electron-spin wave scattering. To analyze the spin wave scattering of conduction electrons in the low temperature range, the zero field resistivity data was fitted in different ways. The resistivity of a magnetic material in the ordered state can be described by the following relations [24]

$$\rho_{metal} \sim T^{(2p+4)/n} \qquad (6) ,$$

$$\rho_{alloy} \sim T^{(2p+3)/n} \qquad (7)$$

where p and n are the exponents and are defined by the form factor and dispersion relation as given below

$$E_k = AK^p \qquad (8),$$

$$F_k = BK^n \qquad (9)$$

For a simple antiferromagnet the value of n is 1 and for a ferromagnet it is 2. The form factor becomes constant for p = 0 for an isotropic exchange interaction. This yields the result that the resistivity of a simple isotropic antiferromagnet due to spin wave excitations is described as $\rho_{metal} \sim T^4$ for metals and $\rho_{alloy} \sim T^3$ for alloys. However, a few authors [24-26] have reported that the resistivity of a magnetic material in the ordered state can be generally described as $\rho(T) = \rho_0 + AT^n \exp(\Delta E/k_B T)$, where $\rho_0$ is the residual resistivity, n is the exponent depending on the type of magnetic ordering and it is expected n=3 or 4 for antiferromagnetic materials [24-26]. In the above relation, $\Delta E$ is the value of spin activation energy generally connected with the anisotropy of the material and $k_B$ is the Boltzmann's constant. In the present case, it was found that the fitting in the exponential form, as shown in Figs. 10 (a,b) gave the best results. This fitting has yielded n close to 2, $\rho_0 = 4.3637 \times 10^{-5}$ ohm-cm and $1.44 \times 10^{-5}$ ohm-cm, A= $2.99 \times 10^{-8}$ ohm-cm/$K^2$ and $2.5 \times 10^{-8}$ ohm-cm/$K^2$, $\Delta E$= 5.7 K and 2.9 K for GdSnGe and TbSnGe respectively. The $\Delta E$ values are comparable to those of some rare earth dodecaborides reported by Gabani *et al.*[27], who fitted the same equation for antiferromagnetic materials. Low temperature fit in ErSnGe was not possible due to the limited data on account of low $T_N$.

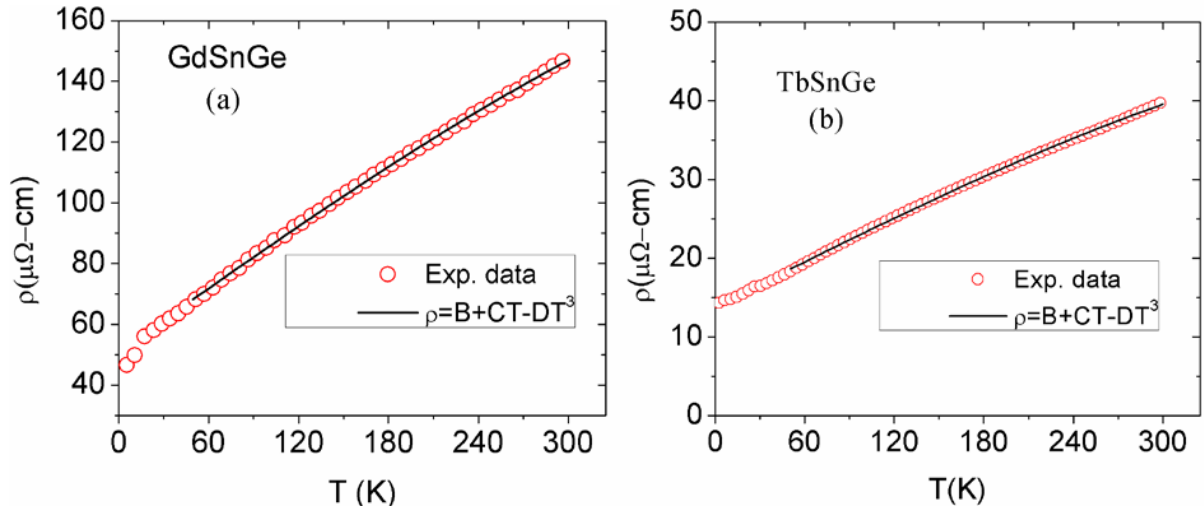

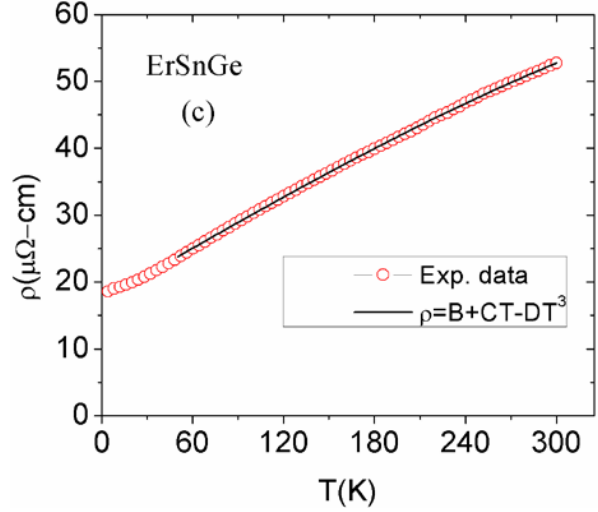

Fig. 11. Fit of equation $\rho = B + CT - DT^3$ to high temperature zero field resistivity data for (a) GdSnGe (b) TbSnGe and (c) ErSnGe.

In metals and alloys, the resistivity at high temperatures is determined by the s-d scattering [28] and has the general behavior given by $\rho(T) = B + CT - DT^3$. In the present case, a fit to the above equation is possible, as shown in Fig. 11(a-c). The values of coefficient B are $5.09 \times 10^{-5}$, $1.38 \times 10^{-5}$ and $1.72 \times 10^{-5}$ ohm-cm, the values of C are $3.48 \times 10^{-7}$, $9.51 \times 10^{-8}$ and $1.3 \times 10^{-7}$ ohm-cm/K and the values of D are $3.16 \times 10^{-13}$, $1.04 \times 10^{-13}$ and $1.38 \times 10^{-13}$ ohm-cm/K$^3$ for GdSnGe, TbSnGe and ErSnGe respectively. The fact that these fits are good implies that the resistivity in the paramagnetic regime is dictated by the s-d scattering.

It can also be noted from Fig. 9 that there is no significant effect of field on the resistivity. All the compounds show very small but positive magnetoresistance (MR) which is expected for antiferromagnetic compounds. There is a sign change of MR at about 235 K in ErSnGe, which is not understood at present.

## IV. Conclusions

In conclusion, all the compounds in RSnGe (R=Gd,Tb,Er) system are antiferromagnetic at low temperatures with varying degrees of coupling strengths. It is seen that Sn/Ge ratio plays a significant role in determining the strength of the antiferromagnetic interaction. Variation in the lattice parameters would be the main reason in determining the strength of the RKKY interaction, which determines the ordering temperature ($T_N$). By comparing the magnetization, heat capacity and resistivity data, it is clear that TbSnGe compound shows some ferromagnetic coupling of moments below 12 K and is strongly antiferromagnetic above that temperature. Among these compounds ErSnGe shows considerable magnetocaloric effect, with a magnetic entropy change of 9.5 J/kg K for a field change of 50 kOe. All the compounds show metallic behavior and small but positive MR. The resistivity in the low and high temperature regimes has been analyzed by fitting to theoretical models.

## Acknowledgments

S. B. G. thanks the C.S.I.R., New Delhi for granting a fellowship. The authors acknowledge the help rendered by D. Buddhikot in carrying out the resistivity measurements.

## References

[1] P. H. Tobash, J. J. Meyers, G. D. Filippo, S. Bobev, F. Ronning, J. D. Thompson, and John L. Sarrao, Chem. Mater., **20**, (2008) 2151–2159.

[2] A. Gil, D. Kaczorowski, B. Penc, A. Hoser, A. Szytula, J. Solid State Chemistry, **184** (2011) 227.

[3] A. Gil, M. Hoffmann, B. Penc, A. Szytula, J. Alloys and Compd., 320 (2001) 29.

[4] A Szytula, D. Kaczorowski, L. Gondek, J. Czub and K. Nenkov, Acta Physica Polonica A, **117** (2010) 586.

[5] Kajuko Sekizawa, J. Phys. Soc. Jap., **21** (1966) 1137.

[6] A. Gil, B.Penc, S.Baran, A.Hoser, A.Szytula, J. Solid State Chemistry, **184** (2011) 1631.

[7] Pramod Kumar, Niraj K. Singh, K. G. Suresh, and A. K. Nigam, Phy. Rev. B, **77** (2008), 184411.

[8] Niraj K. Singh, K. G. Suresh, R. Nirmala, A. K. Nigam, and S. K. Malik, J. Appl. Phys., **99** (2006) 08K904.

[9] Niraj K. Singh, K. G. Suresh, R. Nirmala, A. K. Nigam, and S. K. Malik , J. Appl. Phys., **101** (2007) 093904.

[10] Niraj K. Singh, S. Agarwal, K. G. Suresh R. Nirmala, A. K. Nigam, and S. K. Malik, Phy. Rev. B, **72** (2005) 014452.

[11] E.S.R. Gopal, *Specific Heat at Low Temperatures* (Plenum, New York, 1976).

[12] J. S. Lee, Phys. Stat. Sol.(b) **241**(2004) 1765.

[13] V.K. Pecharsky and K.A. Gschneidner, Jr., J.Appl. Phys., **86** (1999) 565.

[14] E.V. Sampathkumaran, I. Das, R. Rawat, and S. Majumdar, Appl. Phys. Lett., **77** (2000) 418.

[15] P.J. von Ranke,V.K. Pecharsky and K.A. Gschneidner, Jr., Phys. Rev. B, 58 (1998) 12110.

[16] M Klimczak and E. Talik, J. Phys.: Conference Series, **200** (2010) 092009.

[17] N.K. Sun, Y.B. Li, D. Li, Q. Zhang, J. Du, D.K. Xiong, W.S. Zhang, S. Ma, J.J. Liu, Z.D. Zhang, J. alloys and compd., **429** (2007) 29.

[18] R. Malik and E. V. Sampathkumaran, Phys. Rev. B, **58** (1998) 9178.

[19] Kausik Sengupta, S Rayaprol and E.V. Sampathkumaran, J. Phys.: Condens. Matter, **16** (2004) L495-L498.

[20] I. Das, E. V. Sampathkumaran and R. Vijayaraghavan, Phys. Rev. B, **44** (1991) 159.

[21] N. V. Volkenshtein, V.P. Dyakina, V.E. Startsev, Phys. Stat. Sol. (b), **57** (1973) 9.

[22] Naushad Ali and S.B. Woods, Phys. Letters, **104A** (1984) 212.

[23] I. Mannari, Prog. Theor. Phys., **22** (1959) 335.

[24] N. Rivier and A.E. Mensah, Physica B, **91** (1977) 85.

[25] K.N.R. Taylor, M.I. Darby, Physics of Rare Earth Solids, Chapman & Hall, London 1972, p. 208.

[26] J.M. Fournier, E. Gratz, in: K.A. Gschneidner Jr., L. Eyring, G.H. Lander, G.R. Choppin (Eds.), Handbook on the Physics and Chemistry of Rare Earths, Vol. 17, North- Holland, Amsterdam, 1993, p. 423.

[27] S. Gabani, I. Batko, K. Flachbart, T. Herrmannsdorfer, R. Konig, Y. Paderno and N. Shitsevalova, J. Magn. Mag. Mats, **207** (1999) 131-136.

[28] K. G. Suresh and K. V. S. Rama Rao, J. Alloys and compounds, **238** (1996) 90-94